\documentclass[reprint,aps,prx,superscriptaddress,showpacs,onecolumn,12pt]{revtex4-1} 

\usepackage[utf8]{inputenc} 
\usepackage{textcomp} 
\usepackage{graphicx}  
\usepackage{amsmath,amssymb}  
\usepackage{bm}  
\usepackage{color}

\begin{document}

\title{Frequency comb generation in quadratic nonlinear media}

\author{Iolanda~Ricciardi}
\author{Simona~Mosca}
\author{Maria~Parisi}
\author{Pasquale~Maddaloni}
\author{Luigi~Santamaria}
\affiliation{CNR-INO, Istituto Nazionale di Ottica, Via Campi Flegrei 34, 80078 Pozzuoli (NA), Italy}
\author{Paolo~De~Natale}
\affiliation{CNR-INO, Istituto Nazionale di Ottica, Largo E. Fermi 6, 50125 Firenze, Italy}
\author{Maurizio~De~Rosa}
\email[Corresponding author: ]{maurizio.derosa@ino.it}
\affiliation{CNR-INO, Istituto Nazionale di Ottica, Via Campi Flegrei 34, 80078 Pozzuoli (NA), Italy}

\begin{abstract}

We experimentally demonstrate and theoretically explain the onset of optical frequency combs in a simple cavity-enhanced second-harmonic-generation system, exploiting second-order nonlinear interactions. 
Two combs are simultaneously generated around the fundamental pump frequency, with a spectral bandwidth up to about 10~nm, and its second harmonic. We observe different regimes of generation, depending on the phase-matching condition for second-harmonic-generation. 
Moreover, we develop an elemental model which provides a deep physical insight into the observed dynamics. 
Despite the different underlying physical mechanism, the proposed model is remarkably similar to the description of third-order effects in microresonators, revealing a potential variety of new effects to be explored and laying the groundwork for a novel class of highly efficient and versatile frequency comb synthesizers based on second-order nonlinear materials.

\end{abstract}

\maketitle

\section{introduction}
The quest for optical frequency combs (OFCs) was strongly motivated by the need of increasingly precise frequency measurements  and, more recently, of broadband though highly coherent sources. 
Then, OFCs have quickly found new applications beyond frequency metrology and are nowadays routinely used in many laboratories as tools for frequency transfer, precision spectroscopy, astronomical spectral calibration, and generation of low-phase-noise microwave and radio frequency (RF) oscillators~\cite{Newbury:2011dh,Schliesser:2012dn}.
Originally, mode-locked femtosecond lasers were used for producing frequency combs~\cite{Cundiff:2003cn,Maddaloni:2009bg,Adler:2010da}. However, in view of miniaturized photonic tools, comb generation has been demonstrated, in the last years, in continuously-pumped optical microresonators, exploiting the third-order nonlinear susceptibility $\chi^{(3)}$~\cite{DelHaye:2007gi,Savchenkov:2008fw,Foster:2011tr}. 
In such Kerr-combs, the first couple of sidemodes are produced through a degenerate four-wave-mixing (FWM) threshold process, 
 where two pump photons, at frequency $\omega_0/2\pi$, annihilate, creating a pair of signal ($\omega_\text{s}$) and idler ($\omega_\text{i}$) photons, symmetrically placed around the pump, so as to satisfy energy conservation, i.e., $\omega_\text{s} -\omega_0= \omega_0 -\omega_\text{i}$.
The occurrence of self- and cross-phase modulation (SPM and XPM), can compensate the unequal spacing of the cavity modes, due to the group velocity dispersion (GVD) of the material, so that the resonator modes become locally equidistant. 
Successive cascaded FWM processes eventually lead to a uniform broadband frequency comb. 
To date, Kerr-combs have been demonstrated in various geometries, using different materials, and a variety of dynamic regimes and physical features have been observed, stimulating a large number of experimental and theoretical studies~\cite{Kippenberg:2011fc,Chembo:2010cb,Hansson:2013jy,Hansson:2014ie,DelHaye:2014de,Herr:2014ip,Jung:2014jt,Loh:2014ky,Miller:2014dz}.   
It is worth to mention the demonstration and modeling of OFCs in quantum cascade lasers~\cite{Hugi:2012ep,Khurgin:2014hy}.  

Materials with second-order susceptibility, $\chi^{(2)}$,  have been used for transferring and extending otherwise generated OFCs to different spectral regions~\cite{,Adler:2010da,Schliesser:2012dn}. Combs in the near infrared have been transferred in the MIR range by difference frequency generation between a femtosecond comb and a cw source~\cite{Maddaloni:2006ka,Galli:2013cg} or between different teeth of the same comb~\cite{Erny:2007dj,Gambetta:2008fa,Gambetta:2013ci}. 
A more efficient conversion is achieved through optical parametric oscillators synchronously pumped by a femtosecond laser~\cite{Sun:2007wu,Wong:2008fz,Adler:2009ka,Wong:2010gw,Leindecker:2011ch,Keilmann:2012bb}.
Periodically-poled lithium niobate waveguides have been used for spectral broadening of fs fiber lasers generation~\cite{Langrock:2007eg,Phillips:2011is}.
Interestingly, Kerr microresonators can exhibit second-order nonlinearity, whether intrinsic to the material, like AlN~\cite{Jung:2013ju}, or induced by symmetry-breaking the original centrosymmetric structure~\cite{Levy:2011wp,Cazzanelli:2012jz}. 
Even in this case, the effect of the $\chi^{(2)}$ nonlinearity is to frequency up-convert the original $\chi^{(3)}$-comb in the second and third harmonic ranges, but no evidence is reported of a direct intervention of the quadratic nonlinearity in the generation of the fundamental comb~\cite{Jung:2013ju,Miller:2014dz,Jung:2014jt}.

Yet, there is a growing interest in the possibility of direct generation of OFCs, entirely  through $\chi^{(2)}$ interactions, usually more efficient than third-order ones.
Moreover, cascaded $\chi^{(2)}$ processes show a variety of effects typical of $\chi^{(3)}$ materials, like SPM, XPM, FWM, etc.~\cite{Stegeman:1999fe,Saltiel:2005te}, 
which have been exploited in Refs.~\cite{Langrock:2007eg,Phillips:2011is}. 

More recently, Ulvila and coworkers observed frequency comb generation in a singly resonant optical parametric oscillator (OPO) with an additional intracavity crystal, intentionally off-phase-matched for the second-harmonic generation (SHG) of the signal frequency~\cite{Ulvila:2013jv,Ulvila:2014bx}. 
They qualitatively explain their comb generation as a consequence of a Kerr-like SPM occurring in off-phase-matched SHG, where the power propagating in the $\chi^{(2)}$-crystal, initially converted from the fundamental to the second-harmonic wave, after half a coherence length is down-converted back to the fundamental with a phase shift proportional to the fundamental power, finally resulting in an effective optical Kerr effect~\cite{Stegeman:1999fe}. 
However, while Kerr-like SPM can justify the spectral broadening of ultra-fast lasers~\cite{Langrock:2007eg,Phillips:2011is}, for a continuous-wave-pumped crystal it does not necessarily lead to new frequencies, unless FWM is considered as well~\cite{Dudley:2010book}. In addition, off-phase-matched SHG is not essential for the appearance of a comb, as we show in our work.

Here, we experimentally demonstrate frequency comb generation in a continuously-pumped cavity-enhanced SHG system, where multiple, cascaded $\chi^{(2)}$ nonlinear processes enable the onset of broadband $\chi^{(2)}$-comb emission both around the fundamental pump frequency and its second harmonic. 
The observed results are discussed in view of a specially developed dynamical model, which shows a striking resemblance to FWM-based models for Kerr-combs in microresonators~\cite{Chembo:2010cb,Hansson:2013jy}. 
In fact, a properly phase-matched $\chi^{(2)}$ material placed in an optical cavity, singly-resonant for the fundamental frequency, can act either for SHG or OPO, depending on whether it is pumped at the fundamental or a harmonic frequency, respectively. In the former case, the harmonic power generated within the material can exceed the OPO threshold, leading to an internally-pumped cascaded OPO, with steady oscillations of a frequency-symmetric signal/idler (s/i) pair around the fundamental frequency~\cite{Marte:1994wi,Marte:1995vt,Schneider:1997ks,White:1996ts,White:1997ta,Schiller:1997dp,Sorensen:1998vi}.
The occurrence of such an internally-pumped OPO in cavity SHG is usually deleterious for optimal harmonic generation~\cite{Ricciardi:2010kd}. 
Nevertheless, the emergence of unexpected features motivated a series of works which investigated Kerr-like phase shift and sub-harmonic pumped OPO as separately occurring effects~\cite{White:1996ts,White:1997ta,Sorensen:1998vi}. Technical limitations and likely the fact that, at that time, the importance of OFCs was not well understood outside a small circle of people~\cite{Hansch:2006el} prevented an early observation of OFCs in quadratic nonlinear media.  
We show that the cascaded SHG-OPO system displays an even richer dynamics, mimicking typical third-order effects, like those leading to frequency comb generation in $\chi^{(3)}$-nonlinear microresonators.

\begin{figure}[t]
\begin{center}
\includegraphics*[bbllx=20bp,bblly=25bp,bburx=550bp,bbury=560bp,width=0.99\columnwidth]{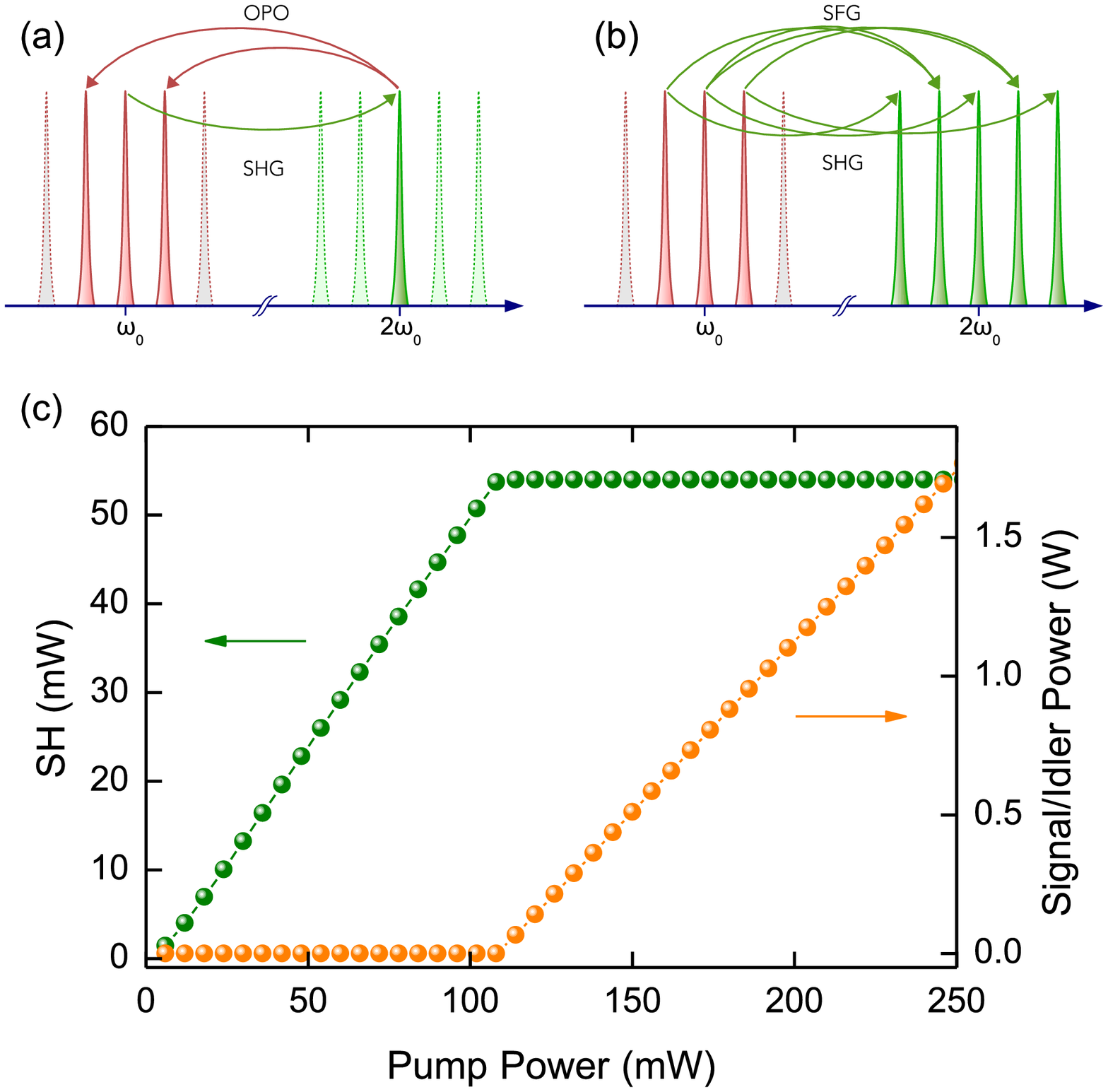}
\caption{Schematic representation of the first steps leading to the formation of a double optical frequency comb in cavity-enhanced second-harmonic generation: (a) second-harmonic generation with cascaded nondegenerate OPO gives rise to multiple sub-harmonic components, (b), which in turn lead to successive, multiple second-harmonic and sum-frequency generations. (c) Second harmonic and intracavity parametric power as a function of the pump power, calculated by numerically solving Eqs.~(\ref{eq:1}) for a set of physical parameters corresponding to the experimental configuration.}
\label{fig:1}
\end{center}
\end{figure}

\section{The dynamical model}
\label{sec2}
A simplified model was previously derived~\cite{Schiller:1997dp}, based on a reduced set of coupled mode equations, considering only the two first processes of frequency doubling and cascaded degenerate OPO: 
a perturbative solution provides, for the three resonating sub-harmonic fields, a set of dynamic equations which displays effective third-order interaction terms; however, not all the relevant terms appear in these equations.
We generalize this model by including the following processes which start at once with OPO onset: generation of the signal and idler second harmonic;  sum frequency of signal (idler) and fundamental  [Fig.~\ref{fig:1}(a) and \ref{fig:1}(b)].
These are all and only processes which lead to a complete, closed three-wave dynamical model for the resonant fields.
Once these processes are considered, the derived dynamic equations are still limited to three sub-harmonic fields, but with new relevant interaction terms. 
Obviously, other possible cascaded processes occur, which indeed lead to the generation of other sidemodes, and possibly to a frequency comb. However, their inclusion in the starting coupled mode equations unnecessarily burdens the present analysis without significantly improving our understanding. 

Then, we derive a complete set of dynamic equations for the sole resonant sub-harmonic fields, i.e., fundamental and parametric fields:
\begin{subequations}
\begin{eqnarray}
\dot{A}_0 &=& -(\gamma + i \Delta_0) A_0 
- 2 g_0 \eta_{0si} A_0^* A_\text{i} A_\text{s} 
\nonumber\\&&
- g_0 (\eta_{00} |A_0|^2+2\eta_\text{0s} |A_\text{s}|^2+2\eta_\text{0i} |A_\text{i}|^2) A_0  
+F_\text{in} 
\label{eq:1-a}
\\
\dot{A}_\text{s} &=& -(\gamma + i \Delta_\text{s}) \, A_\text{s}
- g_0 \eta_\text{00i} \, A_0^2 A^*_\text{i} 
\nonumber\\&&
- g_0 (2 \eta_\text{s0} |A_0|^2 +\eta_\text{ss} |A_\text{s}|^2 + 2\eta_\text{si} |A_\text{i}|^2) A_\text{s}  
\label{eq:1-b}
\\
\dot{A}_\text{i} &=& -(\gamma + i \Delta_\text{i}) \, A_\text{i}  
- g_0  \eta_\text{00s} \, A_0^2 A^*_\text{s}
\nonumber\\&&
- g_0 (2 \eta_\text{i0} |A_0|^2 + 2 \eta_\text{is} |A_\text{s}|^2 + \eta_\text{ii} |A_\text{i}|^2) A_\text{i} 
 \, .
\label{eq:1-c}
\end{eqnarray}
\label{eq:1}
\end{subequations}
Subscripts 0, `s' and `i' indicate fundamental, signal and idler modes, respectively. 
The $A$'s are the normalized electric field amplitudes; $ F_\text{in}$ is the pump amplitude coupled into the cavity; $\gamma$ is the cavity decay constant, assumed to be the same for the three fields; the $\Delta$'s are the cavity detunings of the respective modes; the $\eta$'s are complex nonlinear coupling constants, depending on  the wave-vector mismatches of the considered second-order processes; and $g_0=(\kappa L)^2/2 \tau$ is a common gain factor depending on the crystal length $L$, the second-order coupling strength $\kappa$, and the cavity round-trip time $\tau$ (see Appendix A for a detailed derivation and full mathematical expressions).

Eqs.~(\ref{eq:1}) fully describe, in a compact form, the elemental dynamics of the cavity SHG-OPO system in terms of effective third-order interactions between the three sub-harmonic fields, with the constants $\eta$'s playing the role of third-order complex susceptibilities (it should be noted that here the real part gives the `absorption' component, while the `dispersion' component is the imaginary part, differently from the usual definition of susceptibility). 
The related harmonic fields are {\em fast} variables, which instantaneously---on the cavity round-trip time scale---follow the cavity fields 
(see Appendix A, Eqs.~\ref{eq:SHs}).
However, we note that harmonic fields are not a mere reflection of sub-harmonic fields, but they physically mediate the effective interaction of Eqs.~(\ref{eq:1}), eventually leading to comb formation in both spectral ranges.
The formal analogy between Eqs.~(\ref{eq:1}) and the modal expansion for the Kerr-comb dynamics~\cite{Chembo:2010cb} is remarkable and provides an insightful viewpoint over the dynamical regimes of our system.
A thorough analysis of the steady states of Eqs.~(\ref{eq:1}) and their stability is beyond the scope of the present work. 
Here, we only provide qualitative comments specifically related to what we experimentally observe.

Eqs.~(\ref{eq:1}) predict the onset of a cascaded OPO, above a given input power threshold, and the clamping of the second-harmonic power [Fig.~\ref{fig:1}(c)].  
 More in detail, we focus our attention on some of the interaction terms. 
 Imaginary parts of  terms $|A_l|^2 A_l$, and  $|A_m|^2 A_l$ (with $l,m\in\{0,\text{s},\text{i}\}$ and $l \ne m$), are, respectively, self- and cross-phase modulation terms, producing an effective change of the refractive index which locally compensates the effect of GVD;  
also their real parts play a relevant role, as they determine the frequency distance from the fundamental mode at which a s/i pair oscillates. 
Indeed, of all the s/i pairs that can oscillate, the one with the minimum oscillation threshold prevails, i.e., the one for which the parametric gain exceeds cavity losses first.
In particular, as a s/i pair starts to oscillate, the second harmonics, $2\omega_\text{s/i}$, and sum frequencies, $\omega_\text{s/i}+\omega_{0}$, are generated at once, with an efficiency determined by the respective phase matching conditions. 
The latter sum frequency generations (SFGs) give, in Eqs.~(\ref{eq:1-b}) and (\ref{eq:1-c}), the terms $|A_0|^2 A_\text{s/i}$, whose real part thus represents a nonlinear loss for the respective field $A_\text{s/i}$ (a photon is created in the harmonic region at the expense of a couple of sub-harmonic photons). 
The amount of this loss is generally proportional to the fundamental power $|A_0|^2$, but, more importantly, strongly depends on the value of the corresponding SFG wave-vector mismatch, i.e., on the frequency of the fields (Appendix~B).
As a result, for a given parametric gain and a spectrally equal linear loss, the s/i pair which minimizes the nonlinear losses has the lowest threshold, thus it preferentially oscillates. 
Actually, because of  GVD, the s/i modes of a doubly-resonant OPO generally oscillate with finite detunings $\Delta_\text{s/i}$, resulting in additional effective losses.
Hence, the lowest-threshold parametric pair is determined by a trade-off between cavity dispersion, linear and nonlinear losses, and parametric gain. 
Regarding the terms $|A_\text{i/s}|^2 A_\text{s/i}$, originating from second harmonic of parametric waves, also their real part represents a nonlinear loss; however, at the threshold, they can be neglected in a first approximation, as they are of higher order in the parametric fields.

\begin{figure}[t]
\begin{center}
\includegraphics*[bbllx=0bp,bblly=-1bp,bburx=705bp,bbury=400bp,width=0.99\columnwidth]{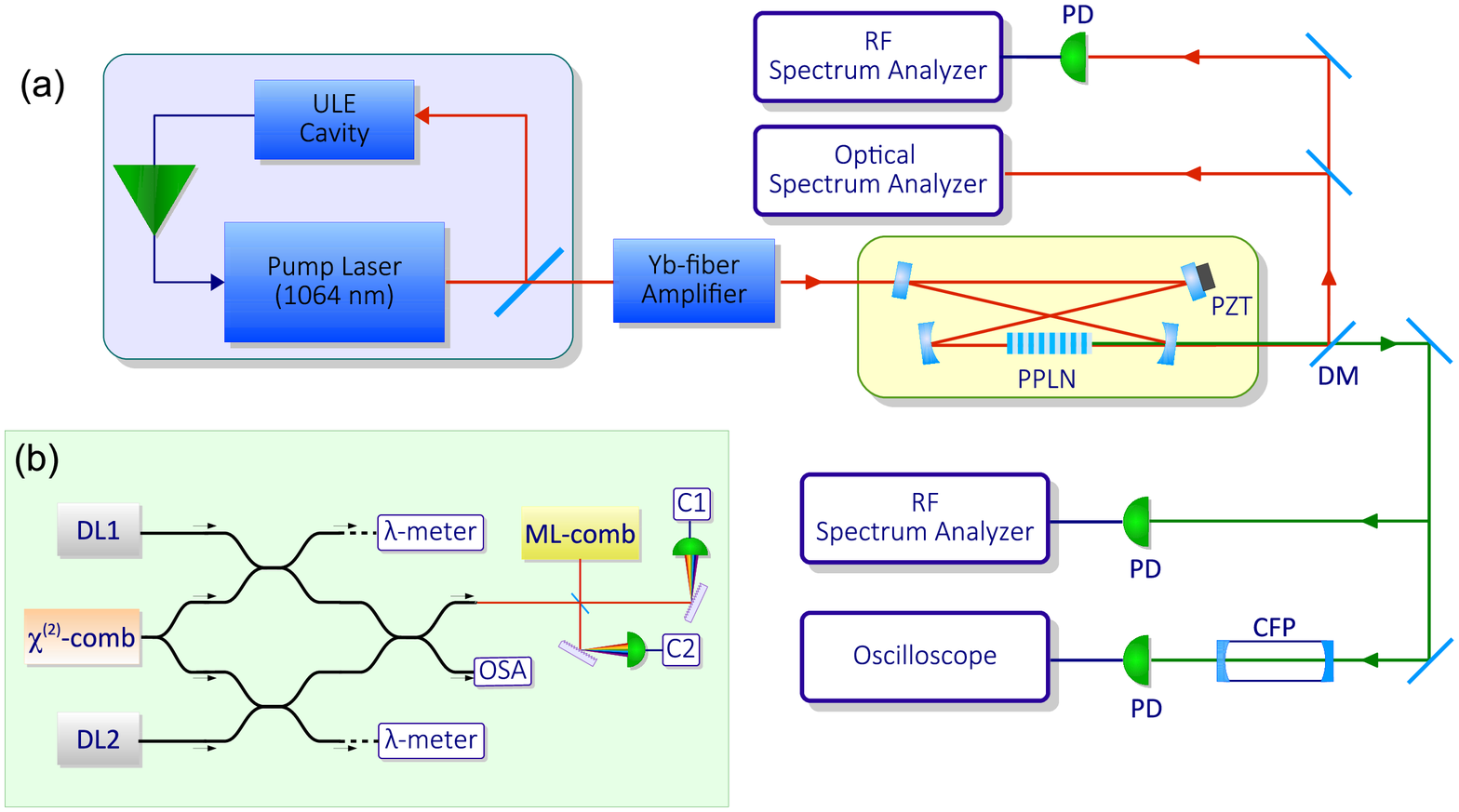}
\caption{Experimental set-up. (a) A four-mirrors travelling-wave cavity, with a periodically-poled  nonlinear lithium niobate (PPLN) crystal inside, is pumped by an amplified cw Nd:YAG laser which is frequency locked to an ultra-low-expansion (ULE) reference cavity. The nonlinear cavity output beams are detected and processed by radio-frequency (RF) analysers, an optical spectrum analyser (OSA), and a confocal Fabry--P\'erot interferometer (CFP)  for spectral analysis in the visible range, not covered by the OSA.
 (b) Frequencies of a couple of $\chi^{(2)}$-comb teeth can be simultaneously measured against an absolutely referenced mode-locked fiber-laser comb (ML-comb), by counting the beat notes of each $\chi^{(2)}$-comb tooth with the corresponding nearest ML-comb tooth. C, frequency counter;  DL, diode laser; DM, dichroic mirror; PZT, piezoelectric actuator.}
\label{fig:setup}
\end{center}
\end{figure}

\section{Experimental setup}
The SHG system is based on a periodically-poled nonlinear crystal, lithium niobate, placed in a travelling-wave optical cavity, resonant for frequencies around the fundamental pump frequency [Fig.~\ref{fig:setup}(a)]. 
The nonlinear cavity consists of two spherical mirrors (100~mm of radius of curvature) and two plane mirrors  in a bow-tie configuration. 
The system is pumped by a cw narrow-linewidth Nd:YAG laser, ($\lambda_0\simeq1064.45$~nm), amplified by an Yb:fibre amplifier (maximum available power, 9~W). 
The pump beam enters the cavity through a 98\%-reflectivity plane coupling mirror, while the remaining mirrors are high-reflection coated ($R>99.98\%$). 
The generated harmonic field exits from the cavity through the first encountered mirror, AR-coated at 532~nm. 
The measured cavity FSR is $493.00 (1)$~MHz, with a cold cavity resonance full width at half maximum of 3~MHz (finesse, 160; Q-factor, $\sim 10^8$).
The crystal, placed between the two curved mirrors, is a 15-mm-long sample of periodically-poled 5\%-MgO-doped lithium niobate, MgO:LiNbO$_3$, with a grating period of $\Lambda=6.96$~$\mu$m. 
The crystal temperature is actively stabilized by a Peltier element driven by an electronic  servo control. The high-reflectivity plane mirror is mounted on a piezoelectric actuator (PZT) for cavity length control. 
The SHG process for the fundamental wavelength of 1064.45~nm is quasi-phase-matched at a crystal temperature $T_{0}=39.5^\circ$C.

The laser is frequency stabilized against an ultra-low-expansion (ULE) cavity by a Pound--Drever--Hall (PDH) locking scheme~\cite{Drever:1983gx} with a residual drift of $\sim1$~Hz/s~\cite{Ricciardi:2013dn}.
A second PDH scheme is implemented to lock the SHG cavity to the pump frequency.
At higher pump powers, photothermal effects strongly distort the PDH signal, preventing active frequency locking. However, the same effects induce a thermal self-locking mechanism~\cite{Carmon:2004us} which enables stable operation, with the laser slightly blue-detuned with respect to the cavity resonance. 
Fundamental and harmonic light beams exiting the cavity are separated by a dichroic mirror and sent to different diagnostic systems:   
two fast photodiodes, whose ac signals are processed by a radio-frequency spectrum analyzer; an optical spectrum analyzer (OSA), with a spectral range from 600 to 1700~nm; 
 a 1-GHz-FSR confocal Fabry--P\'erot cavity, acting as a spectrum analyzer for the visible range not covered by the OSA. 

Frequency measurements of comb teeth separation are made using a commercial OFC synthesizer (Menlo Systems, FC-1500), with a spectral span of 1--2 $\mu$m and mode spacing $f_r=$250~MHz, referenced to the Cs primary standard via the global positioning system[Fig.~\ref{fig:setup}(b)]. 
The beat note $f_b$ between a specific $\chi^{(2)}$-comb tooth and the nearest reference comb tooth is detected and counted, the optical frequency being determined as $\nu=m f_r+f_o+f_b$, where $f_o$ is the reference comb offset frequency. To get the mode number $m$, a diode laser (DL) is frequency tuned within a few tens of MHz from the $\chi^{(2)}$-comb tooth.
In this way, the $m$-th reference comb tooth is the closest to the DL too; then, by measuring the DL wavelength with a 50-MHz-resolution $\lambda$-meter, $m$ can be determined unambiguously. The availability of two DL allows simultaneous measurement of two $\chi^{(2)}$-comb teeth.

\section{Results}
\begin{figure}[t]
\begin{center}
\includegraphics[bbllx=40bp,bblly=40bp,bburx=400bp,bbury=420bp,width=0.99\columnwidth]{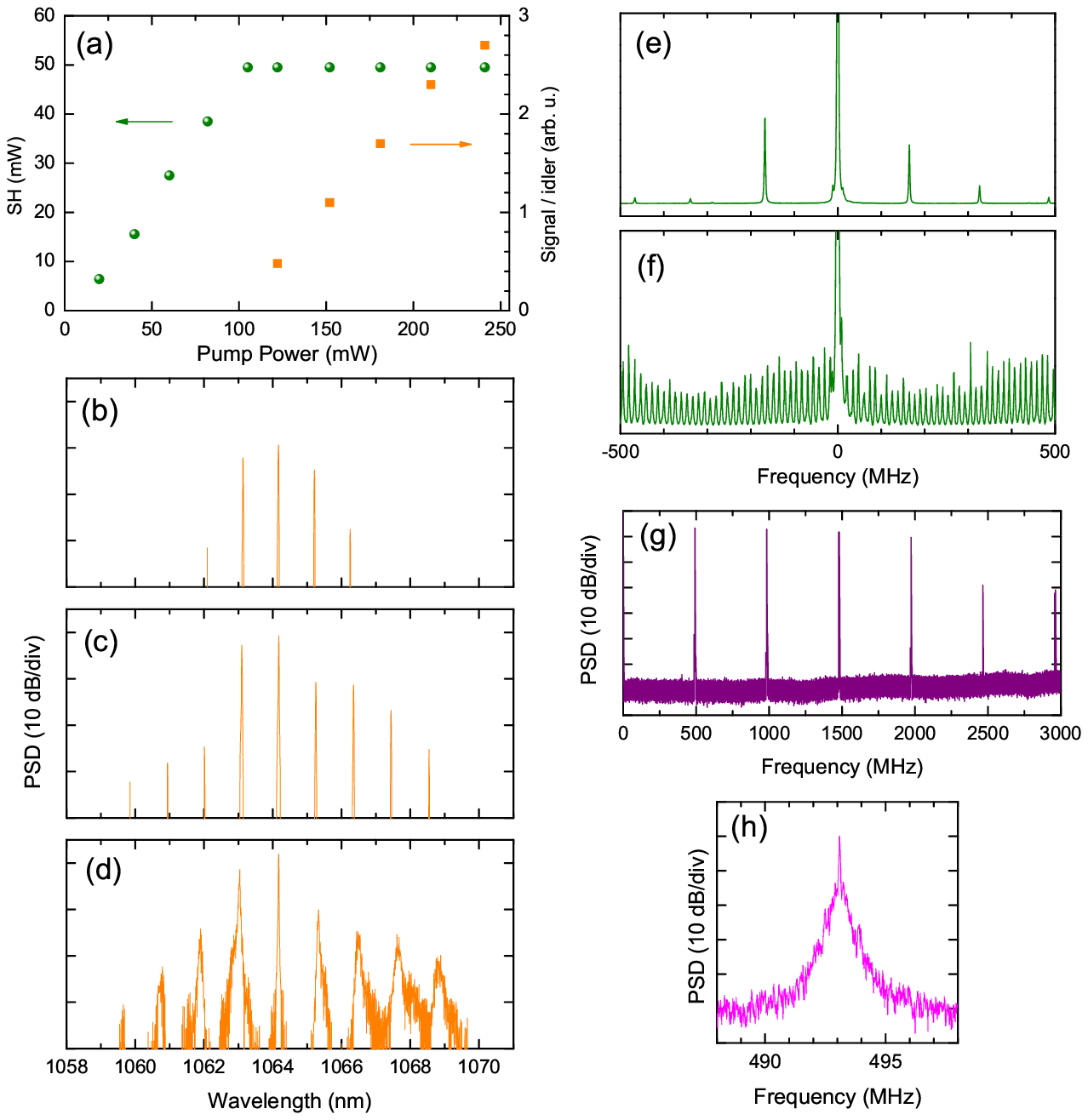}
\caption{Experimental data for quasi-phase-matched SHG: (a) transmitted second-harmonic and parametric power as a function of the input pump laser power; (b)-(d) OSA spectra around the fundamental mode for 170~mW, 2~W, and 9~W of input powers, respectively; (e) and (f),  spectra of the visible CFP corresponding to IR spectra of (b) and (d), respectively; (g) RF spectrum of the IR light output for 9~W of pump power;  (h) detail of the beat note around 493~MHz (RBW=10~kHz, VBW=1~kHz). 
}
\label{fig:comb2}
\end{center}
\end{figure}

When the crystal is quasi-phase matched for SHG, we observe a first regime of pure harmonic generation, at low pump powers, in which the generated second-harmonic power increases with the input pump power, with a SHG efficiency of $\sim50$\% [Fig.~\ref{fig:comb2}(a)]. 
As the laser power exceeds a threshold value of about 100~mW, the second-harmonic power clamps at a constant value, irrespective of the increasing input power, and a s/i pair starts to oscillate at frequencies $\omega_0/2\pi \pm \Delta\nu$  around the fundamental frequency as predicted by Eqs.~(\ref{eq:1}) [Fig.~\ref{fig:comb2}(b)]. 
The frequency separation $\Delta\nu$ corresponds to simultaneous resonant parametric modes which minimize the OPO threshold, determined by a nontrivial trade-off between (linear and nonlinear) losses, GVD, and parametric gain. While the last two processes favour s/i pairs closest to pump, for quasi-phase-matched SHG, the nonlinear loss due to SFG between the fundamental wave and any possible signal or idler is maximum for s/i pairs nearby the pump (see Appendix B). As a result, the lowest-threshold s/i pair rises at a frequency spacing $\Delta\nu$ much larger than the cavity FSR.

Further increasing the pump power above the OPO threshold, additional s/i pairs appear around the fundamental mode, seemingly displaced by multiples of $\Delta\nu$. 
Differently from the first s/i pair, the appearance of successive pairs is thresholdless and can be thought as cascaded nondegenerate FWM between adjacent modes---e.g., the pump mode interacts with each first-order sideband at $\pm\Delta\nu$ generating a new sideband  at $\pm2\Delta\nu$ and amplifying the mode at $\mp\Delta\nu$, and so on, eventually producing a multiple-FSR-spaced frequency comb. 

The  spectral resolution of our optical spectrum analyser  (0.01~nm or $\sim5$~FSR) only enables a rough estimate of the mode separation of the primary comb [Fig.~\ref{fig:comb2}(c)]. A more precise value for the mode separation has been obtained by simultaneously measuring the frequency of two nearby side modes, i.e., first and second order signal (idler) modes [Fig.~\ref{fig:setup}(b)],  having previously measured the cavity-stabilized pump frequency.
This way, we can separately determine the frequency separation between the pump and each first-order sideband, and between first and second order sidebands. 
We finally estimate an equal spacing of $\Delta\nu=288\,406.5$~MHz with a statistical uncertainty of 0.3~MHz, well below the cold-cavity linewidth. In units of FSR the spacing is $\Delta N= (585.00 \pm 0.01)$, an integer multiple within the error. 

When the pump power is further increased, typically $P_\text{in}>5$~W, secondary parametric oscillation and four-wave mixing occur, resulting in the emergence of small (secondary) frequency combs around the primary comb teeth [Fig.~\ref{fig:comb2}(d)].
Similar hierarchical comb formation has been experimentally observed in Kerr-combs~\cite{Papp:2011in,Herr:2012by} and predicted by numerical simulations based on modal expansion~\cite{Chembo:2010cb,Hansson:2014ie}.
At the maximum available power, secondary combs spread towards a continuous spectral distribution, spanning about 10~nm.
The appearance of secondary combs around sidemodes can be understood by considering each primary sidemode as the pump for a secondary elemental threshold process described by Eqs.~(\ref{eq:1}).
Contrary to the quasi-phase-matched fundamental mode,  a primary sidemode is not quasi-phase-matched for SHG and, as a consequence, the nonlinear loss due to its SFG with a possible secondary parametric mode can be minimal close to the primary sidemodes, i.e., a secondary comb is more likely to start closer to the corresponding primary sidemode (Appendix B). 
  
As anticipated, frequency combs are simultaneously generated both around the fundamental pump frequency and its second harmonic. 
Figures~\ref{fig:comb2}(e) and \ref{fig:comb2}(f) report the visible spectra, obtained by the confocal Fabry--P\'erot interferometer, corresponding to IR spectra in Fig.~\ref{fig:comb2}(b) and \ref{fig:comb2}(d), respectively. 
In correspondence with the first s/i pair oscillation, at least five different peaks can be observed. We impute them to fundamental, signal, and idler second harmonic as well as to the sum frequency combinations $\omega_\text{s}+\omega_0$ and $\omega_\text{i}+\omega_0$. We point out that all these processes are not phase matched, except the SHG $\omega_0 \rightarrow 2 \omega_0$.  
The RF spectrum of the IR light output for 9~W of pump power is shown in Fig.~\ref{fig:comb2}(g). The corresponding RF spectrum for the green light is practically identical.
The appearance in the RF spectrum of the intermodal beat notes at the FSR frequency, for both the infrared and visible combs, is a clear evidence of a teeth spacing of one FSR.

\begin{figure}[t]
\begin{center}
\includegraphics*[bbllx=30bp,bblly=35bp,bburx=270bp,bbury=295bp,width=0.99\columnwidth]{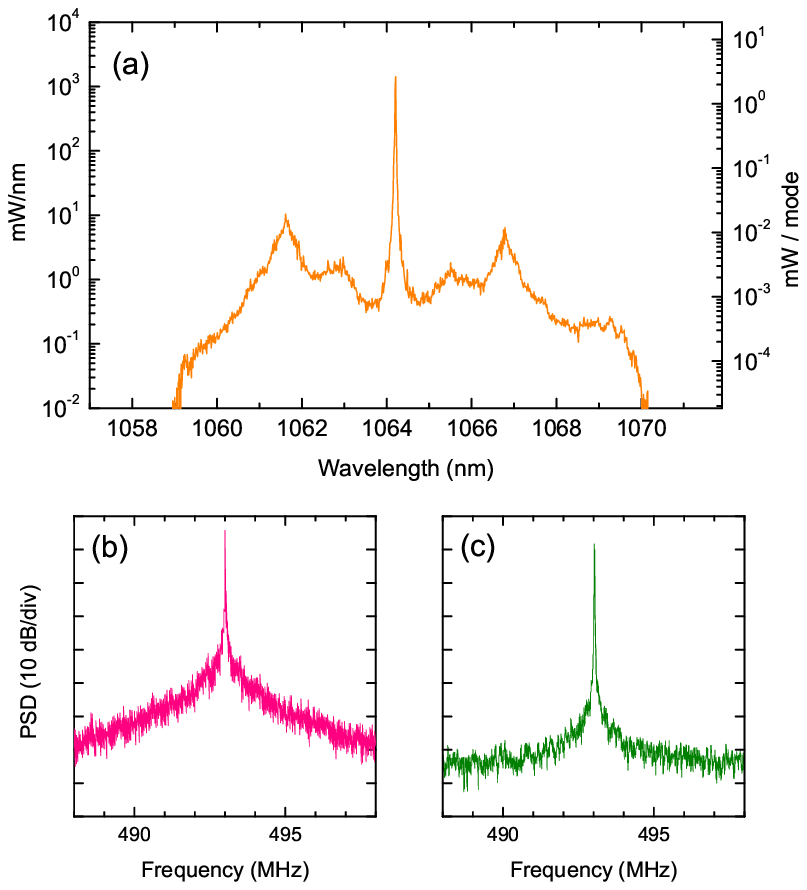}
\caption{Experimental spectra for off-phase-matched SHG. (a) Calibrated optical spectrum of the IR comb emission around the fundamental pump frequency, at the crystal temperature $T=54.2^\circ$C. The integrated power is 50~mW. The scale on the right represents the emitted power per mode. Beat notes around 493 MHz for the IR, (b), and visible, (c), combs (RBW=10~kHz, VBW=1~kHz).}
\label{fig:combOSA+RF}
\end{center}
\end{figure}
Increasing the crystal temperature, the original SHG process becomes positive phase mismatched, namely $\xi_\text{SH0}= k_{2\omega_0} - 2 k_{\omega_0}-K_\text{c}>0$, where $k_{\omega_0}$ and $k_{2\omega_0}$ are the pump and its second-harmonic wave vectors, respectively, and $K_\text{c}=2\pi/\Lambda$ is the crystal periodic-grating wave vector. 
The off-phase matched pump frequency acts as a seed for a comb, similarly to a primary  sidemode in the previous case of quasi-phase-matched SHG. 
In our experiment, we changed the crystal temperature exploring different SHG wave-vector mismatches, for $\xi_\text{SH0} L$ from 0 to $8\,\pi$, with $L$ being the crystal length, limited by the working range of the Peltier servo control. 
Because of the strong photothermal effects, passive thermal-locking is exploited for keeping the cavity nearly-resonant with the pump.

Increasing the pump power, a frequency comb around the fundamental mode emerges and successively broadens up to 10~nm ($\sim5000$ comb teeth). 
As a general trend, the bandwidth of off-phase matched combs increases with the pump power and with the  mismatching temperature.
Figure~\ref{fig:combOSA+RF}(a) shows the power-calibrated optical spectrum of the IR comb emission around the fundamental pump, for a crystal temperature of 54.2$^\circ$C, corresponding to a mismatch $\xi_\text{SH0} L \simeq 8\, \pi$, thus providing an estimate, for the emitted power per mode, of the order of microwatts. 
Figures~\ref{fig:combOSA+RF}(b) and \ref{fig:combOSA+RF}(c) show the beat notes at 493~MHz for the IR and visible combs, confirming the minimal teeth spacing of one FSR. 
Again, when the IR comb emerges around the fundamental wave, a visible comb is also present around its second harmonic.
A comparison can be made between the beat notes of Figs.~\ref{fig:combOSA+RF}(b) and~\ref{fig:comb2}(h):  
both exhibit a linewidth limited by the detection resolution bandwidth, however the former clearly concentrates the power in a narrower spectral range, corresponding to a lower level of intermodal phase noise~\cite{DelHaye:2014de}. 
This suggests a higher degree of correlation between comb teeth in Fig.~\ref{fig:combOSA+RF}(b), with respect to those of Fig.~\ref{fig:comb2}(h), originated by multiple secondary processes. 
The limited spectral coverage of our OSA prevents us to estimate the spectral extension of the harmonic comb.
For negative mismatch conditions, at crystal temperatures $T<T_0$ we are still able to see a multiple-FSR-spaced comb, however the threshold rapidly increases as the temperature decreases and there is no clear evidence of stable closely-spaced combs.

\section{Conclusions}
In conclusion, the emergence of $\chi^{(2)}$-combs in a continuously-pumped external-cavity SHG has been demonstrated and explained according to an elemental dynamical model showing remarkable similarities to Kerr-comb generation in $\chi^{(2)}$ materials. 
Differently from other configurations where $\chi^{(2)}$-nonlinearity is used to replicate or extend an existing frequency comb, our system creates entirely new frequency combs starting from a single-frequency pump. 

Our simple experimental configuration brings to the fore the essential elements which produce the combs, leading to a deeper understanding of the physics through a quantitative and concise theoretical model, of fundamental importance to predict new experiments and design new devices. The resulting formal analogy with third-order comb generation offers a new perspective, stimulating the search of new effects, difficult to envisage on the basis of a purely $\chi^{(2)}$ paradigm, as those observed in Kerr-medium-filled cavities, such as temporal solitons, FWM amplification, intermodal phase coherence and mode-locking, pulsed emission, etc.~\cite{Herr:2014ip,Loh:2014ky}. Furthermore, it confirms a unified approach to frequency combs physics, in whatever way they are generated, and a common theoretical playground for people from different areas. 
We believe that such a model also provides a comprehensive frame for possible generation of OFCs  in other $\chi^{(2)}$-nonlinear systems as well, like those reported in Refs.~\cite{Ulvila:2013jv,Ulvila:2014bx}.

A $\chi^{(2)}$-comb has several advantages with respect to Kerr-combs based on $\chi^{(3)}$ materials, exploiting the intrinsically higher efficiency of $\chi^{(2)}$ processes, combined with the ability of spectrally tailoring the nonlinear efficiency of the material~\cite{Suchowski:2013ks}.
Phase-matching plays a role similar to dispersion in $\chi^{(3)}$ resonators, allowing to change from normal to anomalous ``dispersive'' regimes by varying the phase-matching  condition. 
A thorough analysis of these new phenomena and a generalization of the theoretical model are required for optimal design of new, more efficient frequency comb synthesizers, with lower threshold, larger bandwidth, as well as full frequency stabilization, possibly as small-size, integrated photonic devices~\cite{Savchenkov:2007kn,Furst:2010co,Furst:2010bt,Caspani:2011eo,Cazzanelli:2012jz,Miller:2014dz}.
In principle, $\chi^{(2)}$-combs can be realized all over the transparency range of the nonlinear material, 
a spectral versatility of great importance for the expanding field of direct comb spectroscopy~\cite{Adler:2010da,Galli:2014dt}.
Furthermore, the simultaneous occurrence of octave-distant combs provides a useful metrological link  between two spectral regions without the need for a full octave-wide comb. 
Finally, investigation of quantum properties in $\chi^{(2)}$-based combs is of great importance  as well, in view of the emerging use of multiple correlated photon pairs for multiplexed quantum communication protocols~\cite{Reimer:2014ev}.

\section*{Aknowledgements}
We wish to acknowledge fruitful discussions with J.-J. Zondy. We thank G. Notariale for  technical support. M.D.R. thanks M.~Cossu and her \'equipe for their invaluable cares. 
This work has been partly supported by the Italian ``Ministero dell’Istruzione, dell’Universit\`a e della Ricerca'' (Progetto Premiale QUANTOM---Quantum Opto-Mechanics).

\begin{table*}[t]
\caption{Coupled mode equations for the propagation through the nonlinear crystal, including all the possible $\chi^{(2)}$ up-conversion processes originating from the sub-harmonic fields, $A_0$, $A_\text{s}$, and $A_\text{i}$, being $B_0$ and $B_\text{s/i}$ the pump and signal/idler second harmonic, respectively, and $B_\text{s0/i0}$ the sum frequency of signal/idler and the fundamental mode. }
\label{tab:sistemone}
\begin{equation*}
\tiny
\arraycolsep=2.5pt\def\arraystretch{2.4}
\left.\begin{array}{rcccccc}
\hline
\hline
& \omega_0 \xrightarrow{\rm SHG} 2\omega_0
& 2\omega_0 \xrightarrow{\rm OPO} \omega_\text{s}, \omega_\text{i}
& \omega_\text{s} \xrightarrow{\rm SHG} 2\omega_\text{s}
& \omega_0, \omega_\text{s}  \xrightarrow{\rm SFG}  \omega_\text{s0}
& \omega_\text{i} \xrightarrow{\rm SHG} 2\omega_\text{i}
& \omega_0, \omega_\text{i}  \xrightarrow{\rm SFG}  \omega_\text{i0} 
\\ \hline
\displaystyle \frac{dA_0}{dz}=
&  -i \kappa \, A_0^* B_0 \, e^{-i \xi_\text{SH0} z}  
&  
&  
&  -i \kappa \, A_\text{s}^* B_\text{s0} \, e^{-i \xi_\text{s0}z}
&  
&  -i \kappa \, A_\text{i}^* B_\text{i0} \, e^{-i \xi_\text{i0}z} 
\\ 
\displaystyle \frac{dB_0}{dz}=
&  -i \frac{\kappa}{2} \, A_0^2 \, e^{+i \xi_\text{SH0} z} 
&  -i  \kappa \, A_\text{s} A_\text{i} \, e^{+i \xi_\text{OPO} z} 
&
&
&
&
 \\ 
\displaystyle \frac{dA_\text{s}}{dz}=
&
&  -i \kappa \, A_\text{i}^* B_0 \, e^{-i \xi_\text{OPO} z}
& -i  \kappa \, A_\text{s}^*  B_\text{s} \, e^{-i \xi_\text{SHs}z}
& -i  \kappa \, A_0^*  B_\text{s0} \, e^{-i \xi_\text{s0}z} 
&
& 
\\ 
\displaystyle \frac{dB_\text{s}}{dz}=
&
&
&  -i \frac{\kappa}{2} \, A_\text{s}^2  \, e^{+i \xi_\text{SHs}z}
&
&
& 
\\ 
\displaystyle \frac{dB_\text{s0}}{dz}=
&
&
&
& -i  \kappa \, A_0  A_\text{s}  \, e^{+i \xi_\text{s0}z} 
&
& \\  
\displaystyle \frac{dA_\text{i}}{dz}=
&
&  -i \kappa \, A_\text{s}^* B_0  \, e^{-i \xi_\text{OPO}z}
&
&
&  -i  \kappa \, A_\text{i}^*  B_\text{i}  \, e^{-i \xi_\text{SHi}z}
&  -i  \kappa \, A_0^*  B_\text{i0}  \, e^{-i \xi_\text{i0}z}
\\
\displaystyle \frac{dB_\text{i}}{dz}= 
&
&
& 
&
& -i \frac{\kappa}{2} \, A_\text{i}^2  \, e^{+i \xi_\text{SHi}z}
&
\\
\displaystyle \frac{dB_\text{i0}}{dz}= 
&
&
&
&
&
& -i  \kappa \, A_0  A_\text{i}  \, e^{+i \xi_\text{i0}z}
\\[3pt]
\hline\hline
\end{array}\right.
\end{equation*}
\end{table*}

\appendix
\section{Derivation of effective $\chi^{(3)}$ dynamic equations}
Here we outline the derivation of the system of Eqs.~(\ref{eq:1}). 
We consider collinear plane waves for the interacting modes, whose electric fields propagate along $z$ with slow varying amplitude $E_j(z,t)$,
\begin{equation}
{\cal E}_j(z,t)= \frac{1}{2} E_j(z,t) \, e^{i(\omega_j t -k_j z)} + \text{c.c.} \, ,
\end{equation}
where $\omega_j$ is the angular frequency, $k_j$ is the related wave vector,  explicitly defined in the following.
The coupled mode equations describing the field propagation through a crystal of length $L$ are  displayed in Table~\ref{tab:sistemone}, where the interaction terms are schematically grouped by processes. 
Field amplitudes at a given frequency $\omega_x$, $A$ for sub-harmonic and $B$ for the harmonic range, represent the slow varying electric field amplitudes $E_j$ normalized to $\sqrt{n(\omega_x)/\omega_x}$, where $n$ is the refractive index of the nonlinear crystal.
For each process, a wave vector mismatch must be considered, as
\begin{eqnarray*}
\xi_\text{SH0} &=& k_{2\omega_0} -2k_{\omega_0}
\\
\xi_\text{OPO} &=& k_{2\omega_0} - k_{\omega_\text{s}} - k_{\omega_\text{i}}
\\
\xi_\text{SHs} &=& k_{2\omega_\text{s}} -2k_{\omega_\text{s}}
\\
\xi_\text{s0} &=& k_{\omega_\text{s0}} - k_{\omega_\text{s}} - k_{\omega_0}
\\
\xi_\text{SHi} &=& k_{2\omega_\text{i}} -2k_{\omega_\text{i}}
\\
\xi_\text{i0} &=& k_{\omega_\text{i0}} - k_{\omega_\text{i}} - k_{\omega_0} \, .
\end{eqnarray*}
The coupling constant $\kappa$ can be assumed to be the same for all the processes, as far as the generated sidemodes are nearly degenerate with the fundamental or its second harmonic, i.e., $\omega_0\simeq\omega_\text{s}\simeq\omega_\text{i}$, and $2\omega_0\simeq 2\omega_\text{s} \simeq 2\omega_\text{i}\simeq \omega_\text{s0} \simeq \omega_\text{i0}$, hence~\cite{Yariv:QE}
\begin{equation*}
	\kappa=\frac{d}{c}\sqrt{\frac{2\omega_0^3}{n_1^2 \, n_2}} \, ,
\end{equation*}
with $d$ the effective nonlinear coefficient of the material, $c$ the speed of light, $n_1=n(\omega_0)$, and $n_2=n(2\omega_0)$.

We perturbatively solve the coupled mode equations according to the same procedure  adopted in~\cite{Schiller:1997dp}.
For weakly interacting fields, the equation set of Table~\ref{tab:sistemone} can be easily integrated at the first order, neglecting the spatial variation of the field amplitudes, considering that the nonresonant harmonic fields $B$'s at the crystal input facet, $z=0$, are null, obtaining
\begin{subequations}
\begin{eqnarray}
A_0(z)&=& A_0(0)
\label{eq:1th-a} 
\\
B_0(z)&=& - i \kappa \, [\frac{1}{2} \, A_0^2(0) \, G(\xi_\text{SH0}, z) \nonumber \\
&  & \quad  \quad + A_\text{i}(0) \, A_\text{s}(0) \,  G(\xi_\text{OPO}, z)]
\label{eq:1th-b} 
\\
A_{s/i}(z)&=& A_{s/i}(0)
\label{eq:1th-c}
\\
B_\text{s/i} (z)&=& - \frac{i \kappa}{2} \, A_\text{s/i}^2(0) \,  G(\xi_\text{SHs/SHi}, z)
\label{eq:1th-d}
\\
B_\text{s0/i0}(z)&=& - i \kappa \, A_0(0) \, A_\text{s/i}(0) \,  G(\xi_\text{s0/i0}, z) \, ,
\label{eq:1th-e}
\end{eqnarray}
\label{eq:1th}
\end{subequations}
with
 \begin{equation}
 G(\alpha, z) = \int_0^z e^{i  \alpha  \zeta} \; d\zeta   \, .
\label{eq:gi}
\end{equation}

Substituting the first-order solution in the equation set of Table~\ref{tab:sistemone} and integrating again over the length $L$ of the nonlinear crystal, we finally obtain the second-order expression of the variations of the sub-harmonic fields across the crystal,
\begin{subequations}
\begin{eqnarray}
A_0(L) - A_0(0)	
	&=&
	- \frac{\kappa^2 L^2}{4} \,\eta_{00} \,  |A_0(0)|^2 A_0(0) 
	- \frac{\kappa^2 L^2}{2} \, \eta_\text{0s} \,  |A_\text{s}(0)|^2 A_0(0)	
	\nonumber \\
	& &
	- \frac{\kappa^2 L^2}{2} \, \eta_\text{0i}\,  |A_\text{i}(0)|^2 A_0(0)
	-  \frac{\kappa^2 L^2}{2} \, \eta_\text{0si}  \, A_0^*(0) A_\text{s}(0) A_\text{i}(0)
\label{eq:2nd-a} 
\\
A_\text{s}(L) - A_\text{s}(0) 
	&=& 
	-  \frac{\kappa^2 L^2}{2} \, \eta_\text{s0} \,  |A_0(0)|^2 A_\text{s}(0)
	-  \frac{\kappa^2 L^2}{4} \, \eta_\text{ss} \,  |A_\text{s}(0)|^2 A_\text{s}(0) 
	\nonumber \\
	& &
	-  \frac{\kappa^2 L^2}{2} \, \eta_\text{si}\,  |A_\text{i}(0)|^2 A_\text{s}(0)
	-  \frac{\kappa^2 L^2}{4} \,  \eta_\text{00i}\,  A_0^2(0) A_\text{i}^*(0)
\label{eq:2nd-b}
\\
A_\text{i}(L) - A_\text{i}(0) 
	&=&
	-  \frac{\kappa^2 L^2}{2} \, \eta_\text{i0} \,  |A_0(0)|^2 A_\text{i}(0)
	-  \frac{\kappa^2 L^2}{2} \, \eta_\text{is} \,  |A_\text{s}(0)|^2 A_\text{i}(0) 
	\nonumber \\
	& &
	-  \frac{\kappa^2 L^2}{4} \, \eta_\text{ii} \,  |A_\text{i}(0)|^2 A_\text{i}(0)
	-  \frac{\kappa^2 L^2}{4} \, \eta_\text{00s} \,  A_0^2(0) A_\text{s}^*(0) \, ,
\label{eq:2nd-c}
\end{eqnarray}
\label{eq:2nd}
\end{subequations}
where
\begin{subequations}
\begin{eqnarray}
\eta_{jj} &=& I (\xi_{\text{SH}j},\xi_{\text{SH}j},L) 
\label{eq:etas-1} \\
\eta_{0l} &=&\eta_{l0} = I (\xi_{l0},\xi_{l0},L) 
\label{eq:etas-2} \\
\eta_\text{si} &=& \eta_\text{is} = I (\xi_\text{OPO}, \xi_\text{OPO}, L)
\label{eq:etas-3} \\
\eta_\text{00s} &=& \eta_\text{00i} = I (\xi_\text{OPO}, \xi_\text{SH0}, L) 
\label{eq:etas-4} \\
\eta_\text{0si} &=& I (\xi_\text{SH0}, \xi_\text{OPO}, L)  \, ,
\end{eqnarray}
\label{eq:etas}
\end{subequations}

\noindent
with $j \in \{0,\text{s}, \text{i} \}$, $l \in \{ \text{s}, \text{i} \}$, and
\begin{equation}
I(\alpha,\beta,L) = \frac{2}{L^2} \int_0^L  e^{-i \alpha \zeta} \, G(\beta, \zeta) \; d\zeta \, .
\label{eq:integrale}
\end{equation}

\begin{figure}[t]
\begin{center}
\includegraphics*[bbllx=0bp,bblly=0bp,bburx=310bp,bbury=140bp,width=0.8\columnwidth]{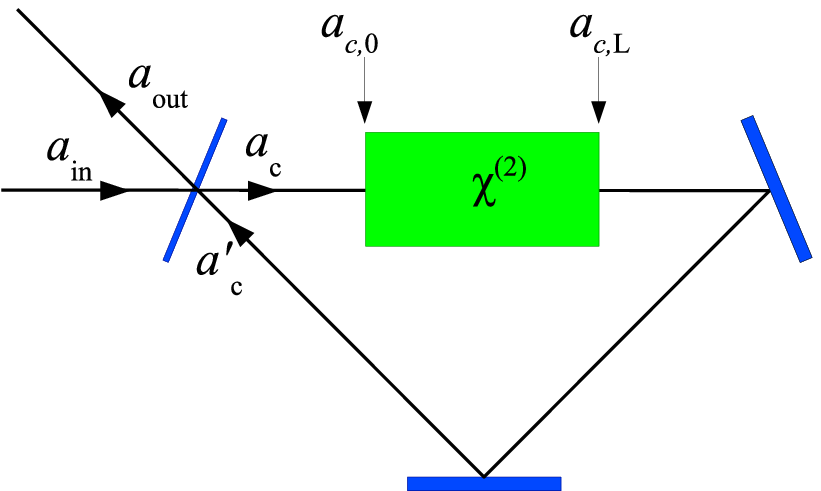}
\caption{Schematic view of a single-ended cavity with a nonlinear medium inside. $a_\text{in}$ and $a_\text{out}$ are the external input and output fields, $a_c$ and $a'_c$ are the cavity fields leaving and impinging on the input mirror, respectively. $a_{c,0}$ and $a_{c,L}$ are the field at the facets of the nonlinear medium.}
\label{fig:consistency}
\end{center}
\end{figure}

Eqs.~(\ref{eq:2nd})  can be used for deriving the rate equations for the cavity dynamics. 
For the sake of simplicity, we consider a lossless single-ended cavity with a nonlinear medium inside, as showed in Fig.~\ref{fig:consistency}. Assuming a slow variation of the resonant cavity field $a_c$  in a round-trip time $\tau$, $\tau \, \dot{a}_\text{c}  \simeq a_\text{c}(t+\tau) - a_\text{c}(t)$, 
the following equation of motion is obtained:
\begin{equation*}
 \dot{a}_\text{c}  = -(\gamma + i \Delta) a_\text{c} + \frac{1}{\tau} [a_{\text{c},L} - a_\text{c,0}]   + \sqrt{2\gamma/\tau} \, a_\text{in} \, ,
\end{equation*}
where $\gamma$ is the power decay rate, $\Delta$ the frequency detuning of the oscillating field with respect to a cavity eigenfrequency, $a_\text{in}$ is a possible input field, and the overdot represents a time derivative. The term in square brackets is the variation of the cavity field due to nonlinear interaction.
We assume that all the sub-harmonic fields experience the same $\gamma$ and $\tau$. 
Inserting the corresponding expression given by Eqs.~(\ref{eq:2nd}), we finally get the  Eqs.~(\ref{eq:1}) describing the dynamics of the three resonating fields.

 As far as the second harmonic and sum frequency fields $B$ are concerned, their dynamics is ``slaved'' to the sub-harmonic fields, i.e. their amplitudes instantaneously follow the sub-harmonic amplitudes, on the time scale of $\tau$, according to
\begin{subequations}
\begin{eqnarray}
B_0(t)&=& - i \kappa \, [\frac{1}{2} \, A_0^2(t) \, G(\xi_\text{SH0}, L)
\nonumber\\
+ A_\text{i}(t) \, A_\text{s}(t) \,  G(\xi_\text{OPO}, L)]
\\
B_\text{s/i} (t)&=& - \frac{i \kappa}{2} \, A_\text{s/i}^2(t) \,  G(\xi_\text{SHs/SHi}, L)
\\
B_\text{s0/i0}(t)&=& - i \kappa \, A_0(0) \, A_\text{s/i}(t) \,  G(\xi_\text{s0/i0}, L) \, ,
\end{eqnarray}
\label{eq:SHs}
\end{subequations}
where $A_\text{0}(t)$, $A_\text{s}(t)$, and $A_\text{i}(t)$, are the cavity field amplitudes given by Eqs.~(\ref{eq:2nd}). We notice that the first order solutions for the $B$ fields, Eqs.~(\ref{eq:1th-b}),(\ref{eq:1th-d}), and (\ref{eq:1th-e}), are valid regardless of the iteration order of the perturbative solution.

\begin{figure}[t]
\begin{center}
\includegraphics*[bbllx=0bp,bblly=0bp,bburx=240bp,bbury=160bp,width=0.9\columnwidth]{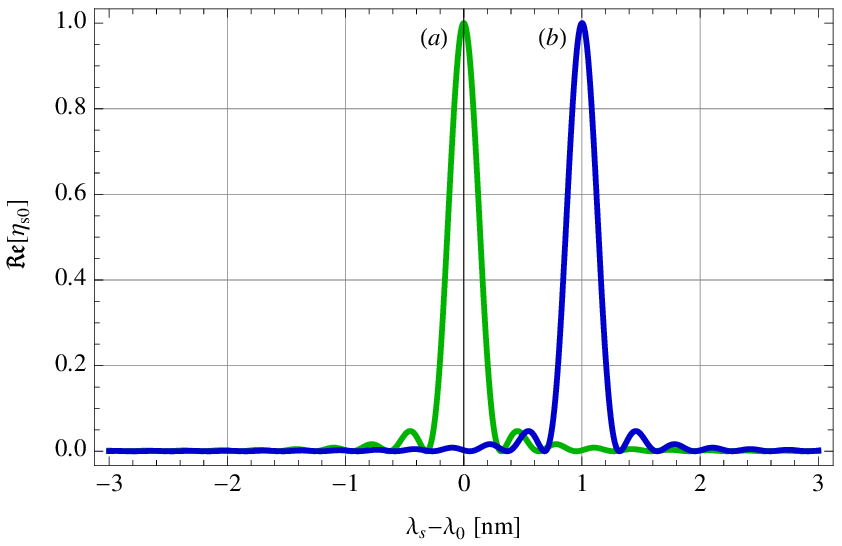}

\caption{Efficiency for SFG of a reference pump mode at $\lambda_0$ with a sidemode at $\lambda_s$ when (a) the pump mode is phase matched for SHG (T=39.5°C) and (b) the pump mode is off-phase matched for SHG (T=45°C).}
\label{fig:real-eta}
\end{center}
\end{figure}

\section{Nonlinear losses at the threshold}
At the threshold of the cascaded OPO, where Eqs.~(\ref{eq:1}) can be linearized with respect to the parametric fields, the real part of  the terms  $2 g_0 \eta_\text{s0} |A_0|^2 A_\text{s}$ and $2 g_0 \eta_\text{i0} |A_0|^2 A_\text{i}$ represents a relevant nonlinear loss for signal and idler, respectively.
Actually, these terms stem from the sum-frequency generation (SFG) processes between the signal (idler) at $\omega_\text{s}$ ($\omega_\text{i}$) and the pump at $\omega_0$.
More in detail, considering the explicit form of the coupling constant $ \eta_\text{s0}$ (analogously for $ \eta_\text{i0}$), by substituting Eqs.~(\ref{eq:gi}) and (\ref{eq:integrale}) in Eq.~(\ref{eq:etas-2}), we obtain the real part,
\begin{equation}
\text{Re} [\eta_\text{s0}] = \text{sinc}^2\left( \frac{\xi_\text{s0}L}{2}\right) \, ,
\label{eq:real-eta}
\end{equation}
which is, in fact, the normalized efficiency of the SFG as a function of the wave vector mismatch $\xi_\text{s0}$.
In Fig.~\ref{fig:real-eta}(a), Eq.~(\ref{eq:real-eta}) is plotted for (a) phase-matched and (b) off-phase matched pump SHG. 
In the former case, highest nonlinear losses occur for signal/idler pairs around the pump frequency, preventing parametric oscillation from starting close to the pump. Conversely, for an off-phase matched pump mode, signal/idler pairs are more likely to oscillate close to the pump, where nonlinear losses can now reach a minimum.
The latter case also applies to secondary comb generation around the teeth of a primary comb, as observed for quasi-phase matched pump SHG.

\begin{figure}[t]
\begin{center}
\includegraphics*[bbllx=40bp,bblly=0bp,bburx=700bp,bbury=520bp,width=0.48\columnwidth]{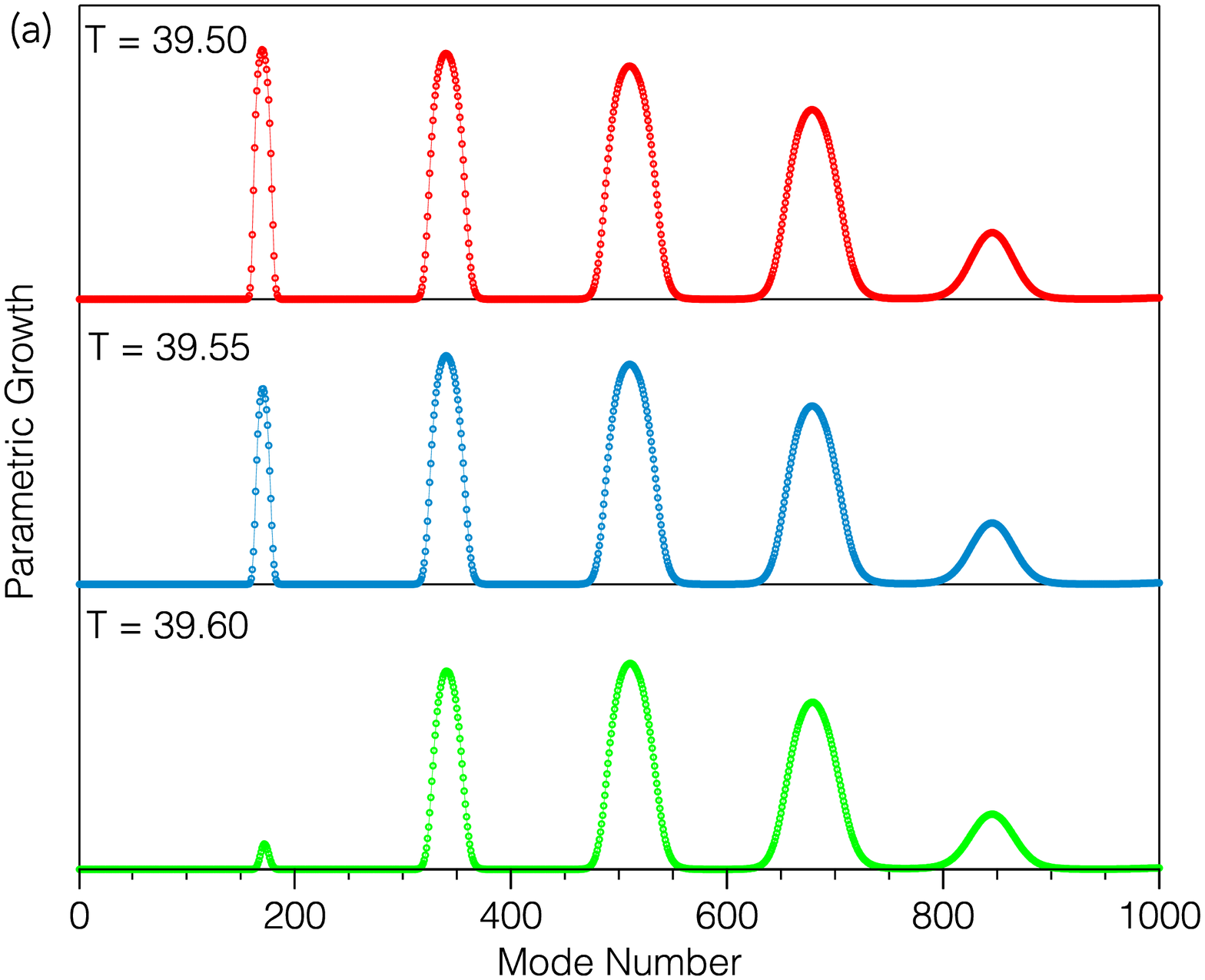}\hfill
\includegraphics*[bbllx=40bp,bblly=0bp,bburx=700bp,bbury=520bp,width=0.48\columnwidth]{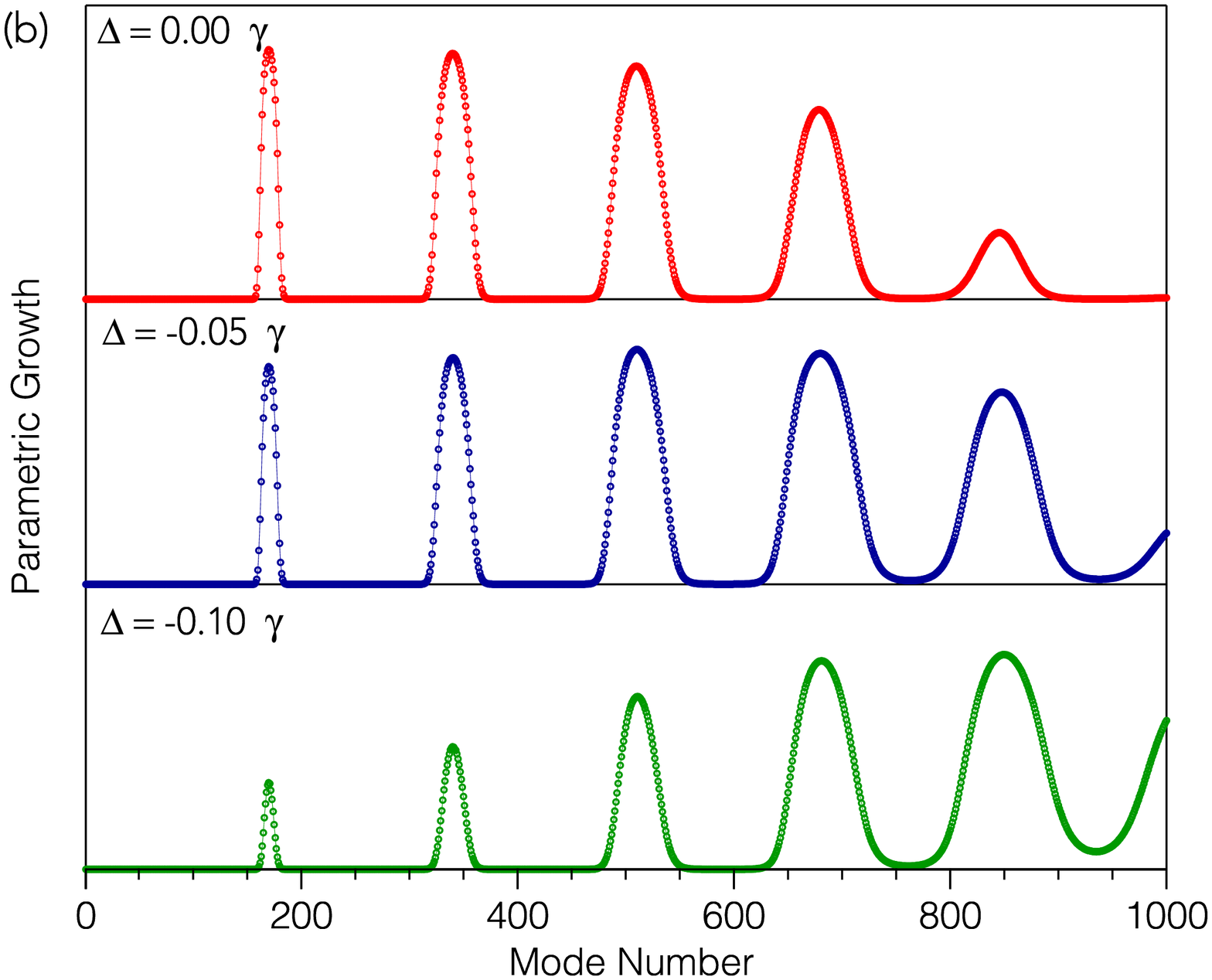}
\caption{Numerical solution of dynamic equations. Parametric growth as a function of the distance of the parametric mode from the pump-resonant mode: (a) for three different crystal temperatures, with a perfectly resonant pump ($\Delta_0=0$); (b) for three different pump detunings, at the phase-matching temperature of 39.5$^\circ$C. The distance is expressed as an integer multiple (mode number) of the cavity FSR and the input power is 110~mW. (The upper curves in panels (a) and (b) are identical.)
}
\label{fig:sim}
\end{center}
\end{figure}

Numerical solutions of Eqs.~(\ref{eq:1}) give a more detailed picture of the final contribution of all the nonlinear losses in determining the power threshold for parametric oscillation and the frequency distance from the pump frequency at which signal and idler appear.
Given a set of values for the input power $F_\text{in}$, pump detuning $\Delta_0$, and the crystal temperature,  Eqs.~(\ref{eq:1}) have been numerically integrated as a function of time until a steady state for parametric oscillation is reached. The procedure has been repeated for each possible frequency-symmetric, nearly-resonant, parametric pair.
Figure \ref{fig:sim} shows a few examples of numerically calculated steady state solutions in the vicinity of the quasi-phase-matching temperature and for small detunings $\Delta_0$ of the pump. For a given input power of 110 mW the parametric power has been calculated as a function of the distance of the parametric mode from the pump frequency, expressed as an integer multiple of the cavity FSR. 
The parametric power is directly related to the parametric gain, hence, the highest parametric power indicates the mode number which is preferred to oscillate. For perfectly phase-matching ($T=39.50^\circ$C) the highest gain is in correspondence with the mode number $N=170$. 
Small deviations from the phase-matching temperature rapidly move the highest gain to the second ($T=39.55^\circ$C, $N=340$) and third ($T=39.60^\circ$C, $N=510$) lobe, as shown in Fig.~\ref{fig:sim}(a). The lobes, with periodic local maxima at multiples of $N=170$, correspond to the first side minima of the SFG curve (a) of Fig.~\ref{fig:real-eta}, in good agreement with the previous qualitative discussion.
In Fig.~\ref{fig:sim}(b) the temperature is kept at $39.5^\circ$C, and the pump detuning is changed. Even in this case, as the detuning increases, the lowest threshold parametric oscillation quickly moves to higher mode numbers.  
The measured teeth spacing agrees within 15\% with the nearest local maxima at $N=510$ of the simulations. 
Our model is limited to three modes; as a matter of fact, additional parametric sidebands start to oscillate as soon as the threshold is surpassed, as in Figure~\ref{fig:comb2}(b). 
Spurious effects, such as etaloning from crystal facets and poling grating could slightly reshape the loss frequency dependence of the ideal model and can explain why in the experiment the parametric sidebands preferentially oscillate at spacings larger than the first side minima. 
Furthermore, our model does not include thermal effects, which can play a significant role in determining the whole evolution of parametric oscillations.
More reliable predictions, including the full evolution of the comb, require a generalization of our model to a large number of oscillating modes.

\bibliography{../X(2)-comb}

\end{document}